\documentclass[journal]{IEEEtran}

\usepackage{todonotes}
\usepackage{amsmath}
\usepackage{placeins}
\usepackage{booktabs}
\usepackage{color}
\usepackage[hidelinks]{hyperref}
\usepackage{enumitem}

\usepackage{amssymb}
\usepackage{mathtools}

\usepackage{booktabs}
\newcommand{\ra}[1]{\renewcommand{\arraystretch}{#1}}

\usepackage{gensymb}
\usepackage{lipsum}

\graphicspath{Figures}

\newcommand{\head}[1]{\textbf{\emph{}}}
\newcommand{\caphead}[1]{\textbf{#1}}

\newcommand{\rev}[1]{{#1}}							

\begin{document}
\title{Gaussian Process Regression for In-situ Capacity Estimation of Lithium-ion Batteries} 
%
%
%

\author{Robert~R.~Richardson,
		Christoph~R.~Birkl,
		Michael~A.~Osborne,
        and~David~A.~Howey,~\IEEEmembership{Member,~IEEE}
\thanks{Authors are with the Department of Engineering Science, University of Oxford, Oxford, UK}%
\thanks{E-mail: robert.richardson@eng.ox.ac.uk, christoph.birkl@eng.ox.ac.uk, mosb@robots.ox.ac.uk, david.howey@eng.ox.ac.uk.}%
}%

\markboth{}%
{Shell \MakeLowercase{\textit{et al.}}: Bare Demo of IEEEtran.cls for IEEE Journals}



\maketitle

\begin{abstract}
Accurate on-board capacity estimation is of critical importance in lithium-ion battery applications. Battery charging/discharging often occurs under a constant current load, and hence voltage vs.\ time measurements under this condition may be accessible in practice. This paper presents a data-driven diagnostic technique, Gaussian Process regression for In-situ Capacity Estimation (GP-ICE), which estimates battery capacity using voltage measurements over short periods of galvanostatic operation.
Unlike previous works, GP-ICE does not rely on interpreting the voltage-time data as Incremental Capacity (IC) or Differential Voltage (DV) curves. This overcomes the need to differentiate the voltage-time data (a process which amplifies measurement noise), and the requirement that the range of voltage measurements encompasses the peaks in the IC/DV curves. GP-ICE is applied to two datasets, consisting of 8 and 20 cells respectively. In each case, within certain voltage ranges, as little as 10 seconds of galvanostatic operation enables capacity estimates with approximately 2--3~\% RMSE.
\end{abstract}

\begin{IEEEkeywords}
Lithium-ion battery, capacity estimation, incremental capacity analysis, diagnostics, Gaussian process regression
\end{IEEEkeywords}

\IEEEpeerreviewmaketitle

\begin{figure}[hbt]
	\centering
	{\includegraphics[width=0.47\textwidth]{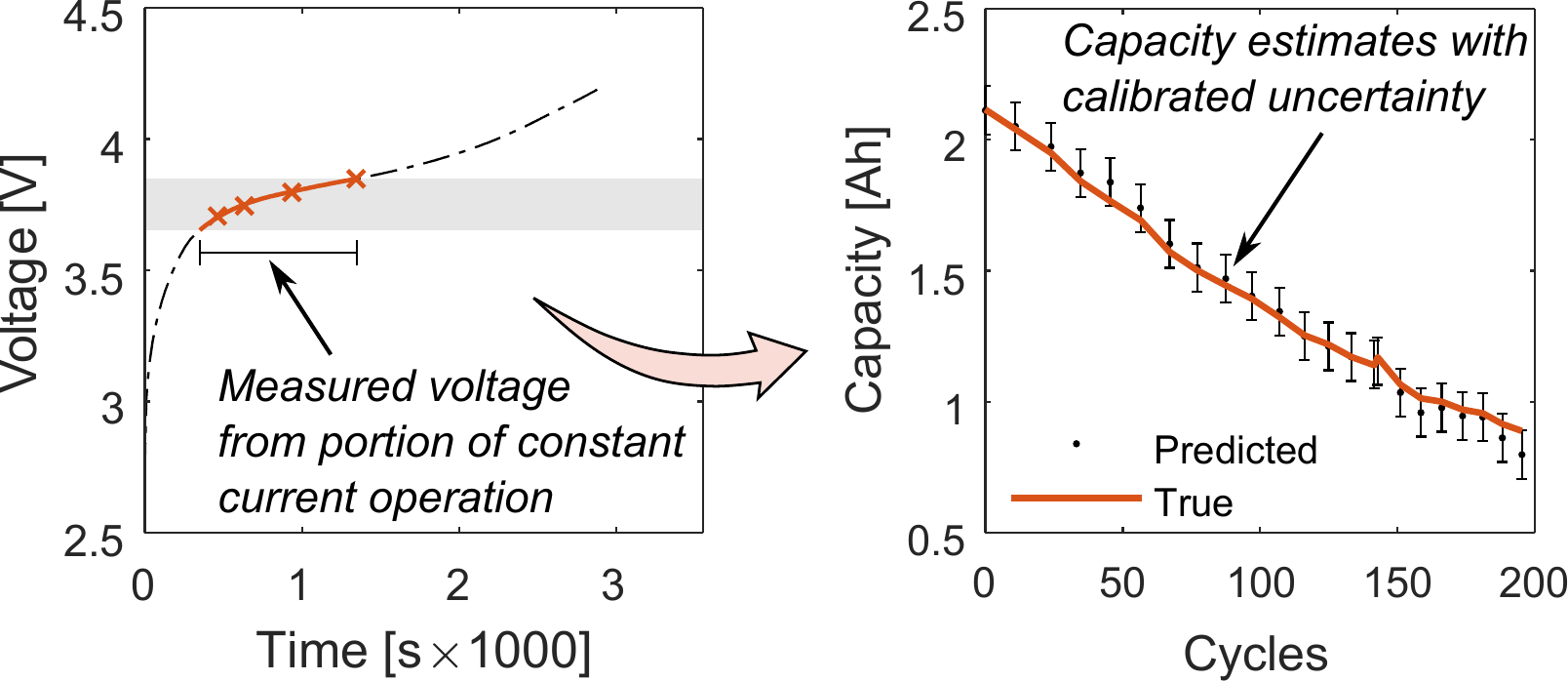}}
	\label{fig:GraphicalAbstract}
\end{figure}

\section{Introduction\label{sec:Introduction}}
%
%
%
%
\IEEEPARstart{L}{ithium-ion} batteries experience capacity fade during use through a complex interplay of physical and chemical processes~\cite{broussely2005main,jung2014understanding}. 
Knowledge of the present battery capacity is necessary to ensure reliable operation and facilitate corrective action when appropriate.
Battery capacity estimates are also an essential input for optimal battery sizing algorithms, for applications such as microgrids \cite{khorramdel2016optimal} and hybrid energy storage systems \cite{shen2014optimization}.
Therefore, accurate online capacity estimation is an important function of the battery management system.


There are several different approaches to capacity estimation~\cite{farmann2015critical,zhang2011review}.
The most common of these involve parameter estimation of battery equivalent circuit models~\cite{plett2004extended,plett2006sigma2,waag2013adaptive,hu2014model} or electrochemical models~\cite{chaturvedi2010algorithms,moura2012pde,prasad2013model,bole2014adaptation}.
These approaches have been successfully applied in many studies; however, they all require the provision of an accurate battery model. Moreover, for high fidelity models, parameter identifiability can be a major challenge~\cite{bizeray2017identifiability}.

Incremental capacity (IC) and differential voltage (DV) analysis have also been used for capacity estimation. These techniques have conventionally been used for detailed cell analyses, such as understanding degradation mechanisms~\cite{dubarry2012synthesize,birkl2017degradation}, however recent studies have considered the use of portions of the IC/DV curve for online capacity estimation \cite{weng2013board,weng2016state,berecibar2016state,berecibar2016online, berecibar2016degradation}.
In particular, Berecibar et al. \cite{berecibar2016online} demonstrated cell capacity estimation using a selection of features of IC/DV curves as inputs.
They demonstrated their approach using three different regression techniques: Linear Regression, Multilayer Perceptrons and Support Vector Machines (SVM), with the latter two methods showing best results.
Although their approach showed good performance, the use of features derived from IC/DV curves as inputs to a regression problem has a number of drawbacks.
Firstly, differentiating the voltage-time data amplifies the noise in the measurement, even when sophisticated smoothing algorithms are employed.
In particular, the magnitude of the peaks were found to be especially sensitive to noise.
Hence, this induces a loss of accuracy in the subsequent regression problem since the inputs are derived from the differentiated data.
Secondly, since the inputs are the values and locations of the peaks, the voltage range must encompass the voltages at which these peaks occur. In some cases, one of these peaks may be located at a high State of Charge (SoC) and another at a low SoC, and hence to identify all the inputs would require covering a large voltage range, and a long measurement duration.
Lastly, the selection of the features is a cumbersome pre-processing step, since these are likely to vary between cells of different chemistries.

The present work overcomes these issues by dispensing with the interpretation of the voltage data as IC or DV curves and instead operating directly on the voltage vs.\ time data itself.
This is achieved by first smoothing the voltage curve using a Savitzky-Golay (SG) filter%
\footnote{Savitztky-Golay filtering is often used when differentiating noisy data; differentiation is not our objective here, however we nonetheless use this filter since it reduces measurement noise, which is advantageous in any case.}%
~\cite{savitzky1964smoothing}, and then using the time values at equispaced voltages as the inputs to the regression problem.
Full details of this procedure, which we term Gaussian Process regression for In-situ Capacity Estimation (GP-ICE), are given in Section~\ref{sec:Method}.
Furthermore, GP-ICE uses Gaussian processes (GPs)~\cite{rasmussen2006gaussian} rather than SVMs or neural networks for the regression step. GPs have previously been used in relatively few studies on battery diagnostics/prognostics~\cite{saha2009prognostics,liu2013prognostics,richardson2017gaussian},
however, they possess a variety of desirable attributes.
Firstly, they are non-parametric\footnote{Support vector machines are, like GPs, non-parametric, but they do not provide confidence estimates in their predictions.}, and hence permit a model expressivity (i.e.\ a number of parameters) that is naturally calibrated to the requirements of the data.
Secondly, GPs are a Bayesian method, and hence handle uncertainty in a principled manner.
An important aspect of diagnostics is not only estimating the capacity values but also expressing the uncertainty associated with these estimates.
Bayesian methods provide a principled approach to dealing with uncertainty, giving rise to credible intervals with probabilistic upper and lower bounds, which are essential for making informed decisions.

The remainder of this article is organised as follows. Section~\ref{sec:Method} describes the novel capacity estimation algorithm, whilst the details of Gaussian process regression are provided in the appendix.
Section~\ref{sec:Datasets} gives details of the two datasets used for validation, and Section~\ref{sec:Results} presents and analyse the results of our method applied to these datasets.
Section~\ref{sec:discussion} discusses the practical applicability of the method, and elaborates on its advantages and disadvantages.

\section{Method\label{sec:Method}}

\subsection{Overview}

An overview of the general methodology is given below. The process is also depicted in Fig.~\ref{fig:exampleTrainingData}, and a detailed flow diagram is included in Fig.~\ref{fig:schematic} at the end of this document. For simplicity, the following description assumes that charging (rather than discharging) data are used, although the procedure is equally applicable in either case.

\subsubsection*{Offline}

Assume we have a database of $N_{C}$ cells. Each cell has been cycled to varying states of health and this cycling may have occurred under varying conditions (e.g. with different C-rates, DoDs, and temperatures).
%
At various stages throughout the life of each cell, a \emph{full} constant-current charge cycle has been applied at a fixed pre-specified current and a fixed pre-specified ambient temperature, and the voltage vs.\ time data from this cycle are recorded. From here on we refer to this data as a Galvanostatic Voltage (GV) curve.
The GV curve is smoothed using a Savitzky-Golay (SG) filter (or any other simple, efficient smoothing algorithm), and the $V$-$t$ data at 1~s intervals are acquired; a subset of these points will be used as the input data for a single sample.
%
Since a full charge/discharge cycle is applied, the capacity of the cell at this C-rate is known and given by $y = \int_{t_{0}}^{t_{\text{end}}}I \, \mathrm{d}t$.
We denote this known capacity as $y$ since it will be the target value for this GV curve in the regression step.
Note that each cell can have a different number of GV curves, and the order of these curves is not important.
Hence, the end result is just a labelled set of training data, consisting of a large set of smoothed GV curves (a subset of which will form the inputs), and an associated set of known cell capacities (the outputs).
The total number of GV curves across all cells is the sample size, $N_{D},$ of the database.

\subsubsection*{Online}
The procedure for estimating the cell capacity using a short online diagnostic test is described next.
Assume we have a cell with an unknown capacity and unknown SoC, and we wish to estimate the capacity.

\begin{figure*}[hbt]
	\centering
	{\includegraphics[width=1\textwidth]{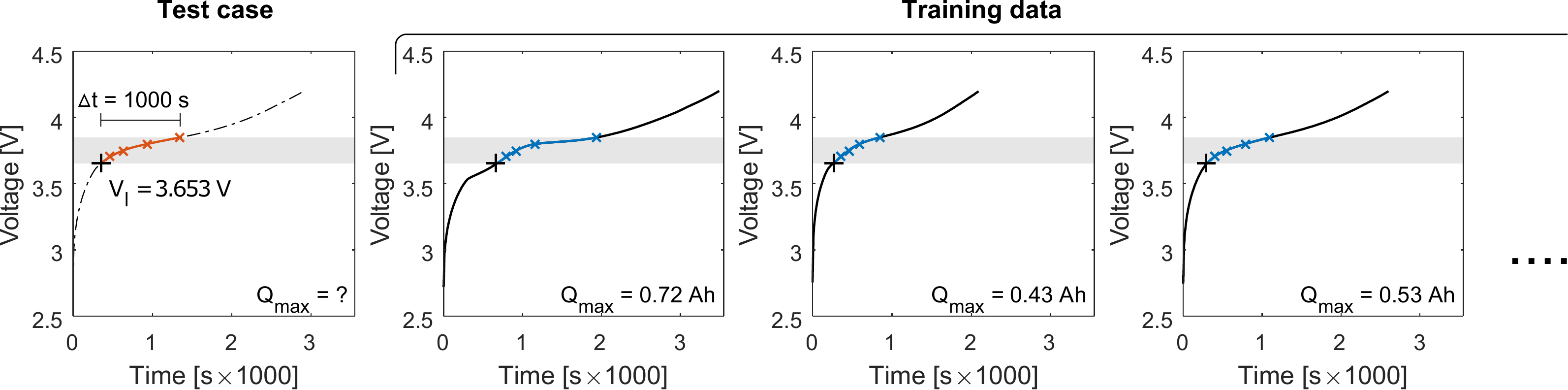}}
	\caption{\caphead{Overview of GP-ICE method.} The time values at $n$ equispaced voltage points between $V_{l}$ and $V_{h}$ are used as inputs to the regression model. The test inputs are shown as red crosses, and the training inputs are shown as blue crosses. The training inputs all have an associated known capacity. The figure shows just three GV curves for training, but in practice the model is trained on several hundred GV curves, obtained from multiple cells at different states of health (see Table~\ref{tab:Data}).}
	\label{fig:exampleTrainingData}
\end{figure*}

\begin{enumerate}
	\item Allow the cell to rest for a sufficient period to minimise electrical/thermal effects from the previous cycle. In practice the minimum duration of this rest will depend on the cell chemistry and the nature of the previous cycle.
	\item Apply the pre-specified constant current for some duration, $\Delta t$, and measure the voltage throughout. In practice $\Delta t$ would be dictated by the duration of time one can afford to take, or the duration of time a device happens to be charged for by the user.
	The voltage range of this test will span from some lower voltage, $V_{l}$, which is the cell voltage when the charge is first applied, to some higher voltage, $V_{h}$, which is the cell voltage at the instant the constant current is removed.
	\item Smooth this voltage vs time data using an SG (or similar) smoothing filter, as before.
	\item Identify the values of the time at $n$ equispaced voltage points between $V_{l}$ and $V_{h}$, and  denote these values by the $n \times 1$ vector $\mathbf{x}^{*}$.
	For example, $n=4$ is chosen, and the voltage spanned from $V_{l}=3.3~$~V to $V_{h}=3.5$~V, then $\mathbf{x}^{*}$ would consist of the time values at  $\mathbf{V} = \{3.35, \, 3.40, \, 3.45, \, 3.50\} \text{V}$,
	i.e.\ $\mathbf{x^{*}}=\mathbf{t}_{\mathbf{V}}$.
	We will later use $\mathbf{x^{*}}$ as the independent variable in the regression model%
	\footnote{Intuitively, the inverse of this procedure (i.e.\ using the voltages sampled at uniformly spaced times as the inputs) might seem to be more logical. However, the former approach is chosen here because using a fixed voltage range prevents the voltage from entering regions where there is no training data. For instance, if a large $\Delta t$ is used in the test case, it might happen that this extends beyond the upper voltage region of the GV curve for a training case with smaller capacity. For example, in Fig.~\ref{fig:exampleTrainingData}, the test case (leftmost subplot) could include up to $\sim2,900$ s (if the entire voltage range was used), whereas this is clearly longer than any possible measurement on the second training case (second subplot from right).}, as shown in Fig.~\ref{fig:exampleTrainingData}.
	\item For each of the GV curves in the offline database, identify the corresponding input vectors, $\mathbf{x}$, given by the time taken to go from the lower voltage to each of the equispaced voltages, i.e. $\mathbf{x}=\mathbf{t}_{\mathbf{V}} - t_{V_{l}}$.
	Since the cell capacities for each GV curve in the offline database are known, each time vector, $\mathbf{x}$, has an associated capacity, which we denote $y$.
	\item Hence, for each GV curve in the training set, there is an input vector $\mathbf{x}$ and an output scalar, $y$. These are used as the inputs and outputs to a GP regression model for predicting the capacity, as described next.
\end{enumerate}

\subsection{Regression}
The goal of a regression problem is to learn the mapping from inputs ${\mathbf{x}}$ to outputs ${y}$, given a labelled training set of input-output pairs $\mathcal{D} = \{(\mathbf{x}_i, y_i)\}_{i=1}^{N_D}$, where $N_D$ is the number of training examples. In the present case, the inputs $\mathbf{x}_i \in \mathbb{R}^{+n}$ are the time vectors for each GV curve, and the outputs ${y}_i \in \mathbb{R}^{+}$ are the corresponding measured capacities, as discussed in the previous section.
The underlying model is assumed to take the form ${y} = f({x}) + \varepsilon$, where $f({x})$ represents a latent function and $\varepsilon \sim \mathcal{N}(0, \Sigma)$ is an independent and identically distributed noise contribution.
The learned model can then be used to make predictions at a test index $\mathbf{x^*}$ (the vector of time values obtained online) for the unknown capacity, ${y^*}$.

In the present work, Gaussian process regression with a Mat\'ern (5/2) kernel function (see Appendix) is used to achieve this mapping.
A full description of the mathematical machinery behind GPs is given in the appendix.
The method was implemented in Matlab using the GPML toolbox~\cite{rasmussen2010gpml}.

A leave-one-out validation scheme was used, whereby each cell is used once as a test set while the data from the remaining cells form the training set.
The performance was evaluated using the root-mean-squared \rev{percentage} error (\rev{RMSPE}) in the capacity estimation, defined as
\begin{equation}
\text{RMSPE}(\hat{y_i}^*, y_i^*) = \sqrt{\frac{1}{N_T}\sum_{i=1}^{N_T}\left(\frac{\hat{y_i}^* - y_i^*}{y_i^*}\right)^2}
\end{equation}
where $\hat{y}^{*}$ is the estimated capacity, $y^{*}$ is the true value, and $N_{T}$ is the total number of test points. \rev{Because percentage errors are normalised, they can be used to compare forecast performance across datasets with different absolute cell capacities, as is the case in this study~\cite{swanson2011mape}.}

To quantify the reliability of the uncertainty estimates, we use the calibration score (CS), defined as the frequency of actual results lying within a given credibility interval. For instance, for a $\pm2\sigma$ credibility interval, the calibration score is defined as:
\begin{equation}
\text{CS}_{2\sigma} = \frac{1}{N_T}\sum_{i=1}^{N_T}\left[\lvert \hat{y_i}^* - y_i^* \rvert < 2\sigma \right].
\label{eq:calib-score}
\end{equation}
For a Gaussian predictive distribution, the interval corresponding to $\pm2\sigma$ is a 95.4\% credibility interval. Hence, the frequency of actual results lying in these intervals should be approximately 0.954: greater or less than this implies that the model is under- or over- confident respectively.

\section{Datasets\label{sec:Datasets}}

Two different datasets are considered in this work: (i) the Oxford dataset, consisting of our own in-house aging experiments and (ii) the NASA dataset, obtained from an open-access repository provided by the NASA Ames Research Centre.
An overview of each dataset is given in Table~\ref{tab:Data}.

\begin{table}
	\centering
	\small
	\ra{1.1}
	\begin{tabular}{lp{0.32\columnwidth}p{0.32\columnwidth}}
		\toprule
		\textbf{Dataset} & \textbf{Oxford} & \textbf{NASA}\\
		\midrule
		Manufacturer & Kokam & LG Chem. \\
		Form factor & Pouch & 18650 \\
		\# cells & 8 & 20 \\
		\# samples & 519 & 842 \\
		$Q$ range (Ah) & 0.74 $\rightarrow$ 0.43 & 2.10 $\rightarrow$ 0.80 \\
		Cycling & All cells cycled with same regime & 5 groups each with different regime \\
		\bottomrule
	\end{tabular}
	\caption{\caphead{Dataset overview.} The `\# samples' column indicates the total number of voltage-time curves, across all cells. The `$Q$ range' column indicates the values of the maximum initial capacity and minimum final capacity respectively, across all cells.}
	\label{tab:Data}
\end{table}

\subsection{Oxford}

The Oxford data was obtained from the Oxford Battery Degradation Dataset\footnote{\hyperlink{https://ora.ox.ac.uk/objects/uuid:03ba4b01-cfed-46d3-9b1a-7d4a7bdf6fac}{https://ora.ox.ac.uk/objects/uuid:03ba4b01-cfed-46d3-9b1a-7d4a7bdf6fac}}~\cite{birkl2017oxford}. This consists of aging experiments applied to 8 commercial Kokam pouch cells of 740 mAh nominal capacity, with graphite negative electrode and lithium cobalt oxide (LCO)/lithium nickel cobalt oxide (NCO) positive electrode.
Cycling was conducted using a Biologic MPG 205 potentiostat, and the cells were housed in a Binder MK53 thermal chamber at a constant ambient temperature of 40 \degree C.

\begin{figure}[hbt]
	\centering
	{\includegraphics[width=0.5\textwidth]{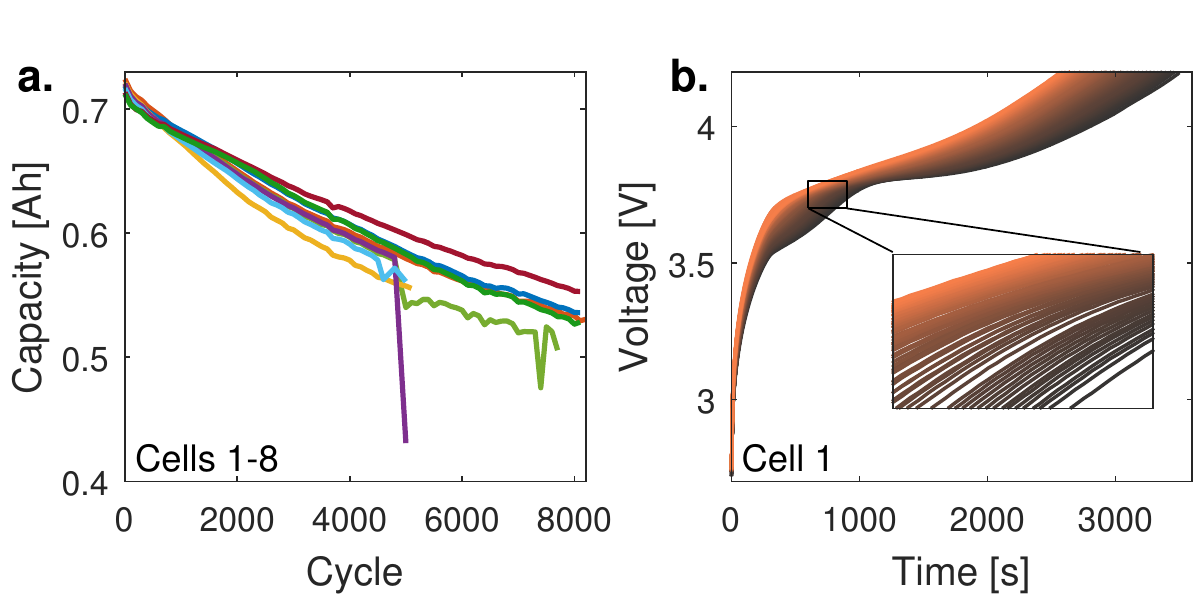}}
	\caption{\caphead{Oxford dataset.} \textbf{a}, Capacity evolution of the tested cells. \textbf{b}, Evolution of the voltage curves for Cell 1 over the life of the cell. The colours range from dark to light as the cycle number increases.}
	\label{fig:dataOxf}
\end{figure}

All 8 cells were cycled by repeatedly discharging using the ARTEMIS urban drive cycle~\cite{andre2009artemis} and recharging at a constant current of 2C. After every 100 cycles, a characterisation test was carried out including a full charge-discharge cycle at 1C -- these were the GV curves for this dataset. Fig.~\ref{fig:dataOxf}b shows the complete set of GV curves for Cell 1 over its entire lifecycle. Similar sets of curves were observed for the other cells. Each of these curves represents a single sample from which the inputs to the regression problem are sampled, as discussed in Section~\ref{sec:Method}. A total of 519 charge curves were measured across all cells (i.e.\ $\sim65$ curves per cell).

The cell capacity was calculated by integrating the 1C charge curves. The calculated capacities for all 8 cells are plotted as a function of cycle number in Fig.~\ref{fig:dataOxf}a.
The end of life (EoL) was deemed to occur if the cell terminal voltage dropped below 0~V during the discharge cycle.
The EoL typically occurred at $\sim$8,000 cycles (Fig.~\ref{fig:dataOxf}a) although one of the cells failed much earlier than this ($\sim$5,000 cycles). Another cell (light green line in Fig.~\ref{fig:dataOxf}a) entered a change of regime around 5,000 cycles where a sudden drop in capacity occurred -- this provides an interesting challenge for the capacity estimation algorithm as discussed in Section~\ref{sec:Results}.

\subsection{NASA\label{sec:Datasets-NASA}}
The NASA dataset was obtained from the NASA Ames Prognostics Center of Excellence Randomized Battery Usage
\href{https://ti.arc.nasa.gov/tech/dash/pcoe/prognostic-data-repository/}{Repository} \cite{bole2014randomized}. The data in this repository was first used in Ref.~\cite{bole2014adaptation} for an investigation into capacity fade under randomized load profiles. The data are randomised in order to better represent practical battery usage.
The tests were conducted with LG Chem.\ 18650 Li-cobalt cells with 2.1 Ah nominal capacity.
The remainder of this subsection describes the cycling and characterisation procedure based on the documentation provided with the downloaded datasets~\cite{bole2014randomized}.

\begin{table}\centering
	\small
	\ra{1.1}
	\begin{tabular}{p{0.49\textwidth}}
		\toprule
		\textbf{Group 1 (Cells 1, 2, 7, 8)}\\
		Repeatedly charged to 4.2V using a randomly selected duration between 0.5 hours and 3 hours, and then discharged to 3.2V using a randomized sequence of discharging currents between 0.5A and 4A. Reference characterisation carried out every 50 cycles.\\
		\textbf{Group 2 (Cells 3-6)}\\
		Same as group 1 except charging cycle is not randomized.\\
		\textbf{Group 3 (Cells 9-12)}\\
		Operated using a sequence of charging/discharging currents between -4.5A and 4.5A. Each loading period lasted 5 minutes. Reference characterisation carried out after 1500 periods (about 5 days).\\
		\textbf{Group 4 (Cells 13-16)}\\
		Repeatedly charged to 4.2V and then discharged to 3.2V using a randomized sequence of discharging currents between 0.5A and 5A. A customized probability distribution designed to be skewed towards selecting higher currents was used to select a new load setpoint every 1 minute during discharging operation.\\
		\textbf{Group 5 (Cells 17-20)}\\
		Same as group 4 except the probability distribution was designed to be skewed towards selecting lower currents.\\
		\bottomrule
	\end{tabular}
	\caption{\caphead{NASA data load profiles.} Each group of cells underwent a different loading procedure. Full details of these procedures are described in the repository documentation~\cite{bole2014randomized}}
	\label{tab:NASA-groups}
\end{table}

\begin{figure*}[t]
	\centering
	{\includegraphics[width=1\textwidth]{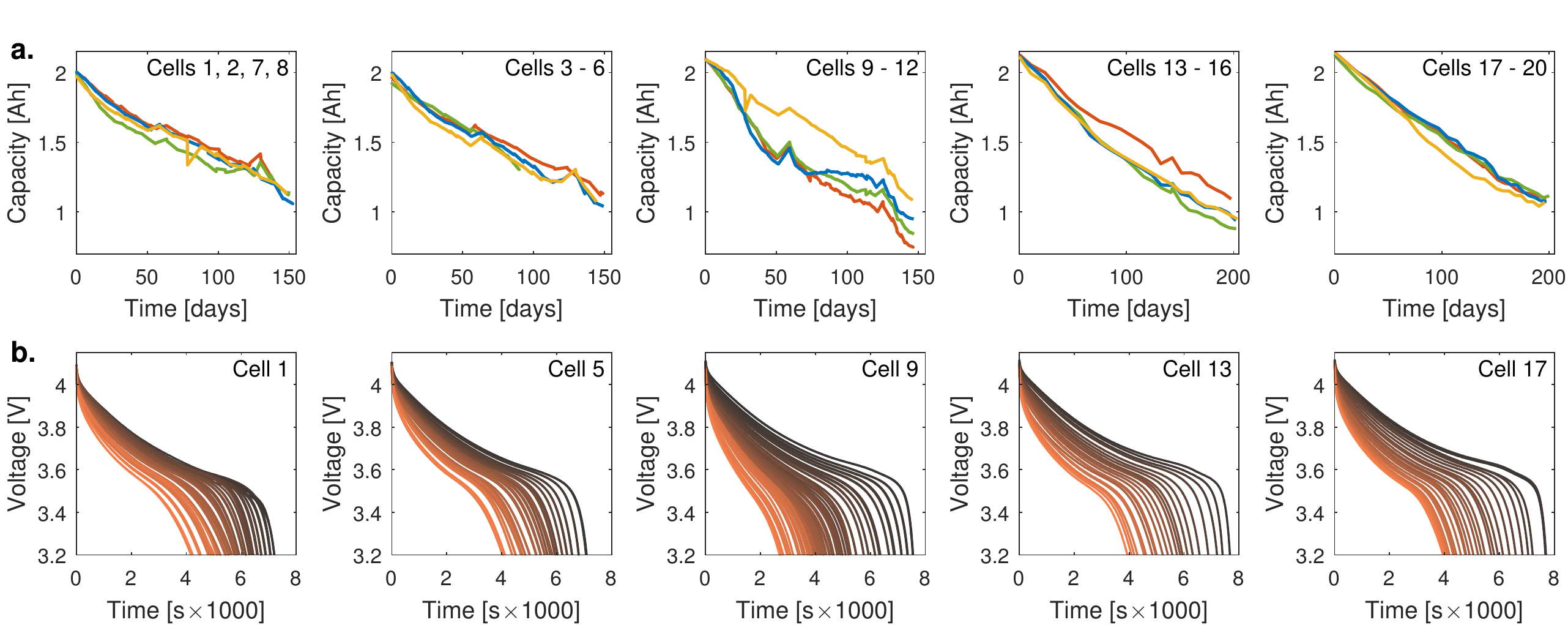}}
	\caption{\caphead{NASA dataset.}  \textbf{a}, Capacity evolution of the 5 groups of tested cells. Each group consists of four cells cycled with similar profiles. \textbf{b}, Evolution of the voltage curves for an exemplary cell from each group. The colours range from dark to light red as the cycle number increases.}
	\label{fig:dataNasa}
\end{figure*}

For this study we used the data from the first 20 cells in the repository, which were all cycled at room temperature throughout the duration of the experiments.
The cells are grouped into 5 groups of 4, with each group undergoing a different randomized loading procedure as described in Table~\ref{tab:NASA-groups}.
In all cases a characterisation test was periodically carried out, whereby a 2A charge-discharge cycle was applied -- the \emph{discharge} curves were used as the GV curves in this case, to demonstrate the applicability of our method using either charge or discharge data.
A total of 842 GV curves were measured across all cells (i.e.\ $\sim42$ curves per cell).

The cell capacity was calculated by integrating the 2A charge curves. The calculated capacities for the cells in all 5 groups are plotted against the cycle count in Fig.~\ref{fig:dataNasa}a. The full set of GV curves for a selected cell from each group is plotted in Fig.~\ref{fig:dataNasa}b, beneath the corresponding capacity plots.
Fig.~\ref{fig:dataNasa} shows that the evolution of the capacity is quite different for each group of cells.
Later results demonstrate that the GP-ICE method is robust in that it provides accurate estimates in spite of this path dependence of the capacity fade.

\section{Results\label{sec:Results}}

\subsection{Oxford dataset}

\begin{figure*}[!t]
	\centering
	{\includegraphics[width=0.99\textwidth]{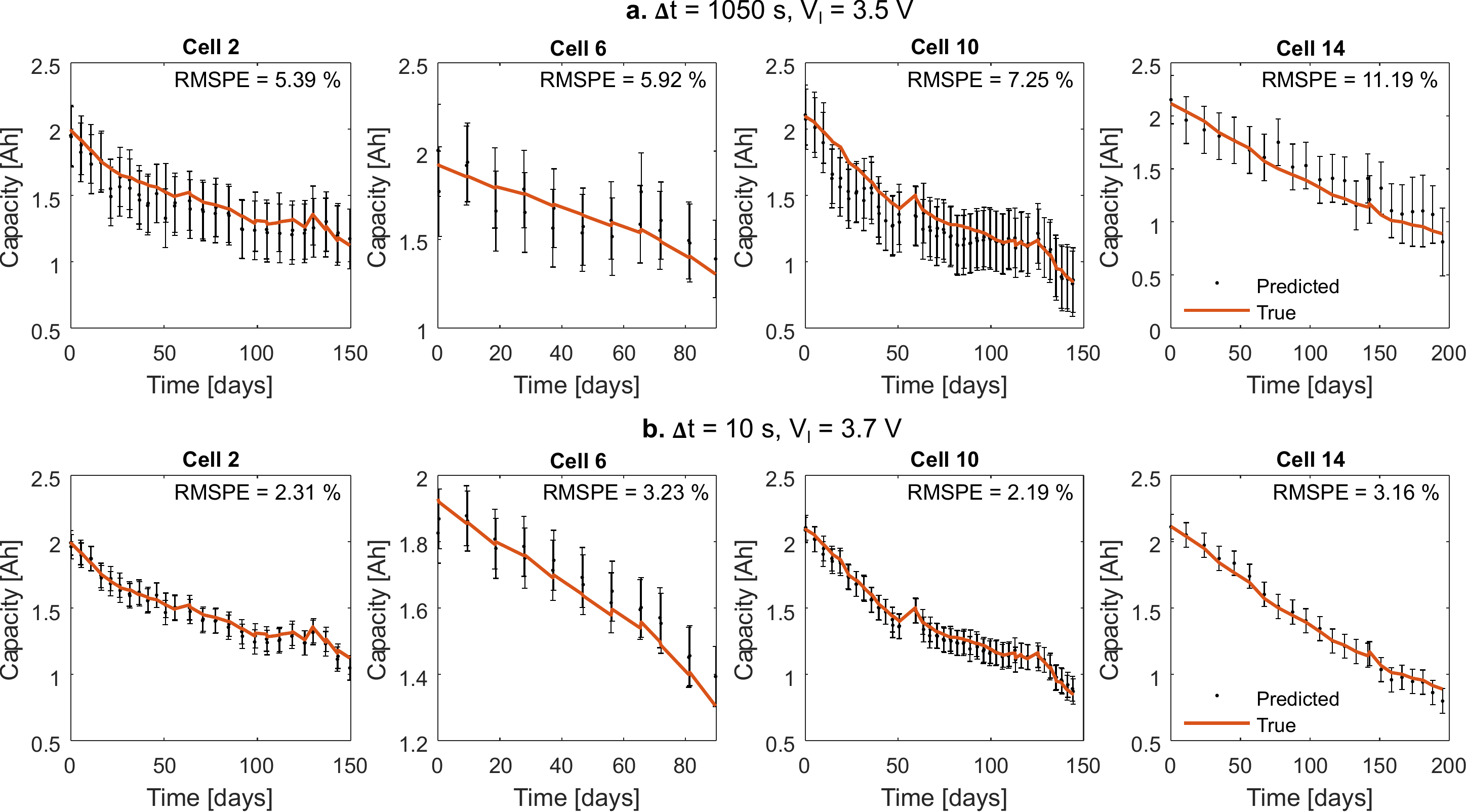}}
	\caption{\caphead{Selected results for the Oxford dataset.} The red lines indicate the measured capacity and the black markers with errorbars indicate the GP-ICE estimates $\pm 2 \sigma$. \textbf{a}, Using a test duration of $\Delta t = 50$ s and starting voltage of $V_{l} = 3.3\text{V}$, \textbf{b}, Using a test duration of $\Delta t = 1050$ s and starting voltage of $V_{l} = 3.5\text{V}$.}
	\label{fig:sampleResultsOxf}
	\vspace{0.4cm}
	\vfill
	{\includegraphics[width=0.49\textwidth]{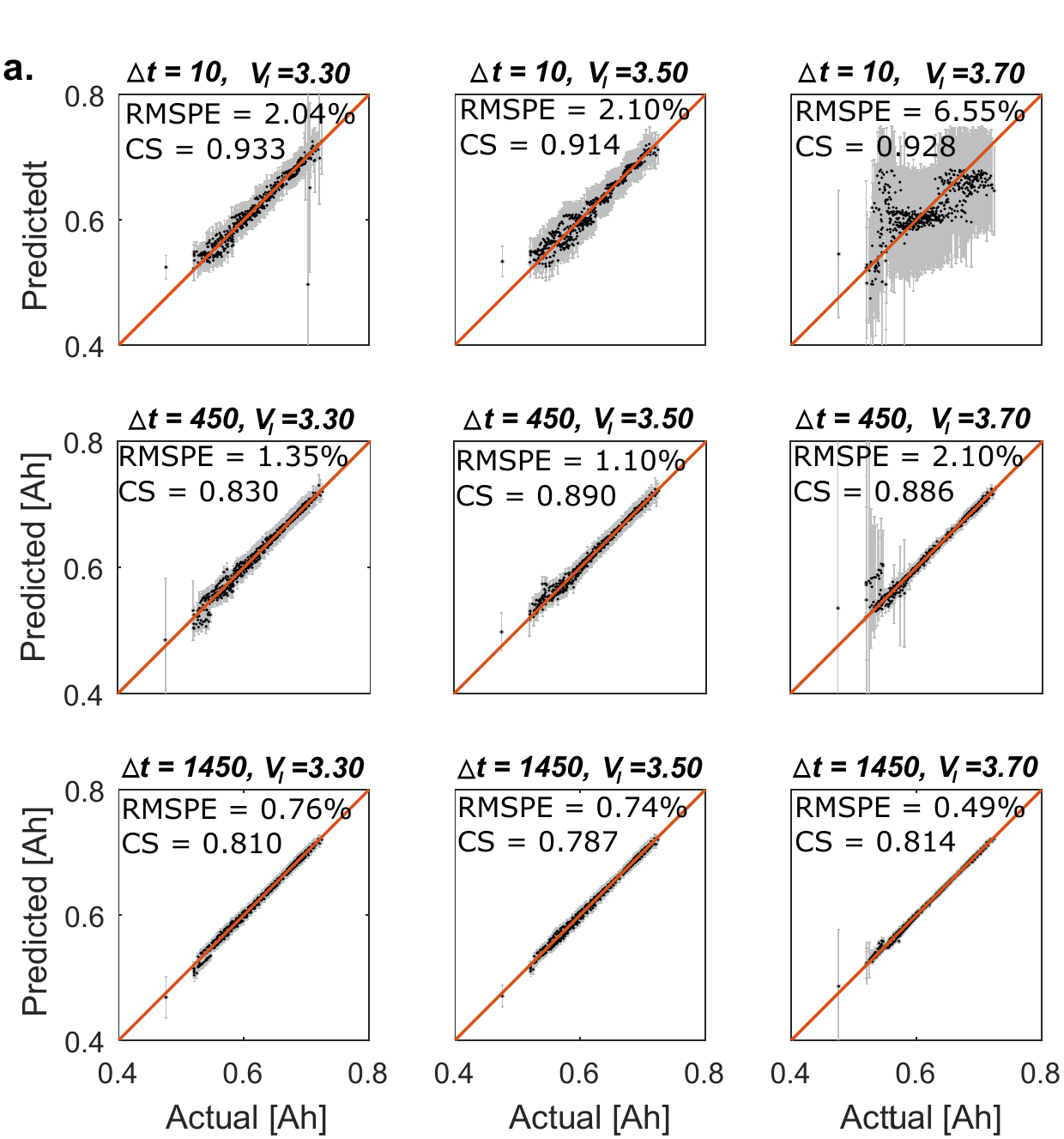}}
	\hfill
	{\includegraphics[width=0.49\textwidth]{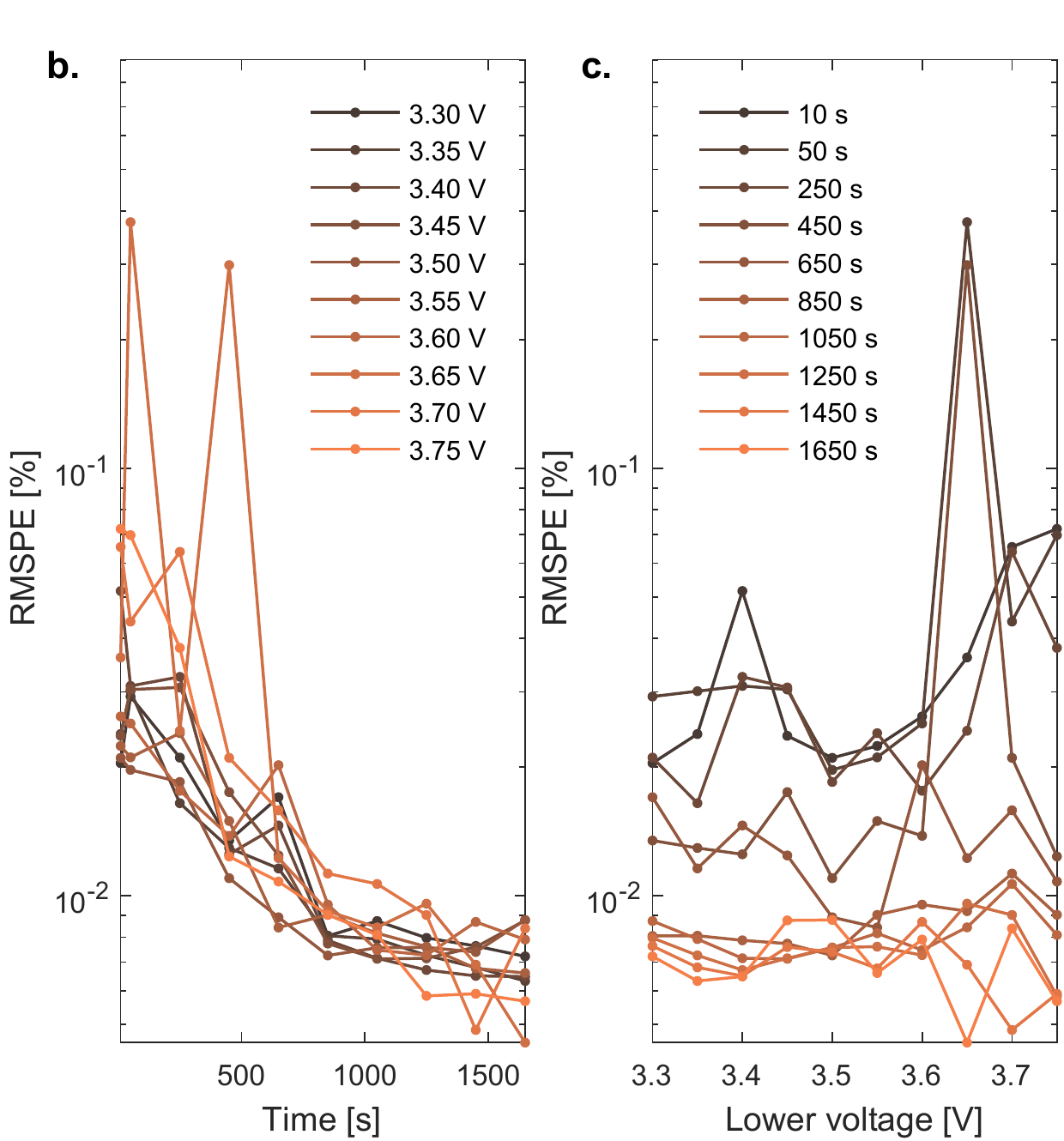}}
	\caption{\caphead{Overall results for the Oxford dataset.} \rev{RMSPE} values are based on the entire dataset with each cell used once as the test set. \textbf{a}, Actual vs.\ predicted capacities for different starting voltages and measurement durations. The red line represents $y^{*} = \hat{y}^{*}$. The closer the datapoints lie to this line, the smaller the difference between the actual and predicted value. The grey lines indicate $\pm 2\sigma$ credibility intervals for each datapoint. The quoted CS values indicate the associated $\pm2\sigma$ calibration score; the closer these scores are to 0.954 the more accurate the uncertainty estimates. \textbf{b}, \rev{RMSPE} vs. measurement duration for different starting voltages. \textbf{c}, \rev{RMSPE} vs.\ starting voltage for different measurement durations. The \rev{RMSPE} clearly decreases with measurement duration but shows relatively little dependence on the starting voltage.}
	\label{fig:resultsOxf}
\end{figure*}

Fig.\ \ref{fig:sampleResultsOxf} shows results for selected cells from the Oxford dataset for two combinations of online measurement duration, $\Delta t$, and lower voltage, $V_{l}$. For each plot, the model is tested on the cell shown and trained on all other cells. Note that for the test set, we do not actually carry out a separate online diagnostic test as described in Section~\ref{sec:Method}; rather the relevant portion of the data was simply selected from the full GV curve, as though it had come from a short diagnostic test.
Fig.\ \ref{fig:sampleResultsOxf}a shows that reasonable performance can be achieved using a relatively short measurement duration of just 50~s. Where the predictions are less accurate, the error bars are quite honest and generally extend to encompass the true values. For instance, Cell 2 exhibits an unusual drop in capacity at $\sim$5000 cycles, a behaviour which is not manifested by any of the other cells (which were used for training in this case).  Hence, the estimates made for Cell 2 after $\sim$5000  cycles are slightly erratic, but their uncertainty is accurately reflected by their correspondingly larger error-bars.
On the other hand, Fig.\ \ref{fig:sampleResultsOxf}b shows that consistently high performance can be achieved if a large $\Delta t$ is used.
The estimates for all cells in this case have an \rev{RMSPE} value below 1\%. Interestingly, the method performs well for Cell 2 even in the regime beyond $\sim$5000 cycles, and expresses high confidence in these estimates.
In practice the provision of such confidence estimates has significant implications. For instance, in an online setting, as capacity measurements are received sequentially from diagnostic tests of varying duration, a Kalman filter~\cite{grewal2011kalman} (or other probabilistic filter) could effectively discount the uncertain measurements and retain the certain ones. This a more robust diagnosis over multiple cycles.

Fig.\ \ref{fig:resultsOxf} shows the overall results, where each cell is used once as the test set.
Fig.\ \ref{fig:resultsOxf}a shows actual vs.\ predicted capacities across all cells for a selection of $\Delta t$ and $V_{l}$ values.
It is apparent that larger $\Delta t$ values (lower rows on the grid of plots in Fig.\ \ref{fig:resultsOxf}a) have higher accuracy, whereas differences in $V_{l}$ (columns of the same grid) have a less consistent effect on the \rev{RMSPE}. This is shown explicitly in Figs.\ \ref{fig:resultsOxf}b and c, which show the overall \rev{RMSPE} values plotted against $\Delta t$ and $V_{l}$ respectively. For all starting voltages there is a clear decreasing trend in \rev{RMSPE} as $\Delta t$ is increased, as would be expected.

For the measurement duration of $\Delta t = 1450$ s (the bottom row of the grid of plots in Fig.\ \ref{fig:resultsOxf}a), the capacity is accurately estimated even at extreme values. For instance, the lone data-point at just under 0.5~Ah lies very close to the red line despite not having other nearby training examples from which to learn. One of the advantages of Bayesian methods such as GPs over deterministic methods is that they can generalise better from relatively small datasets such as the one used here by properly expressing their uncertainty about the underlying model.

On the other hand, when smaller $\Delta t$ values are used (such as the middle and upper rows of plots) this outlier is over-estimated. However, in most cases where the estimates are inaccurate, the error bars are correspondingly larger, hence accurately conveying the model's uncertainty (as indicated by the grey error bars generally crossing the red line in Fig.~\ref{fig:resultsOxf}a).
To evaluate the accuracy of the uncertainty estimates we calculate the calibration score (Eq.~\ref{eq:calib-score}) for two different intervals: CS$_{0.67\sigma}$ and CS$_{2\sigma}$, corresponding to 50\% and 95.4\% credibility intervals, respectively.
The average calibration scores for the model across all combinations of $\Delta t$ and $V_{l}$ are CS$_{0.67\sigma} = 0.432$ and CS$_{2\sigma} = 0.849$; the CS$_{2\sigma}$ values for specific combinations of $\Delta t$ and $V_{l}$ are also quoted within the subplots in Fig.~\ref{fig:resultsOxf}a. These values are slightly less than the corresponding true credibility intervals, which indicates that the model is slightly over-confident in its estimates. This is most likely due to the fact that the model assumes that the inputs are uncorrelated, when in fact they come from a GV curve with sequential structure.
However, these uncertainties are still quite reasonable, especially in comparison to non-probabilistic approaches (such as the previously used neural networks or SVMs \cite{berecibar2016online}), which implicitly assign equal credibility to all estimates.

\begin{figure*}[!t]
	\centering
	{\includegraphics[width=0.99\textwidth]{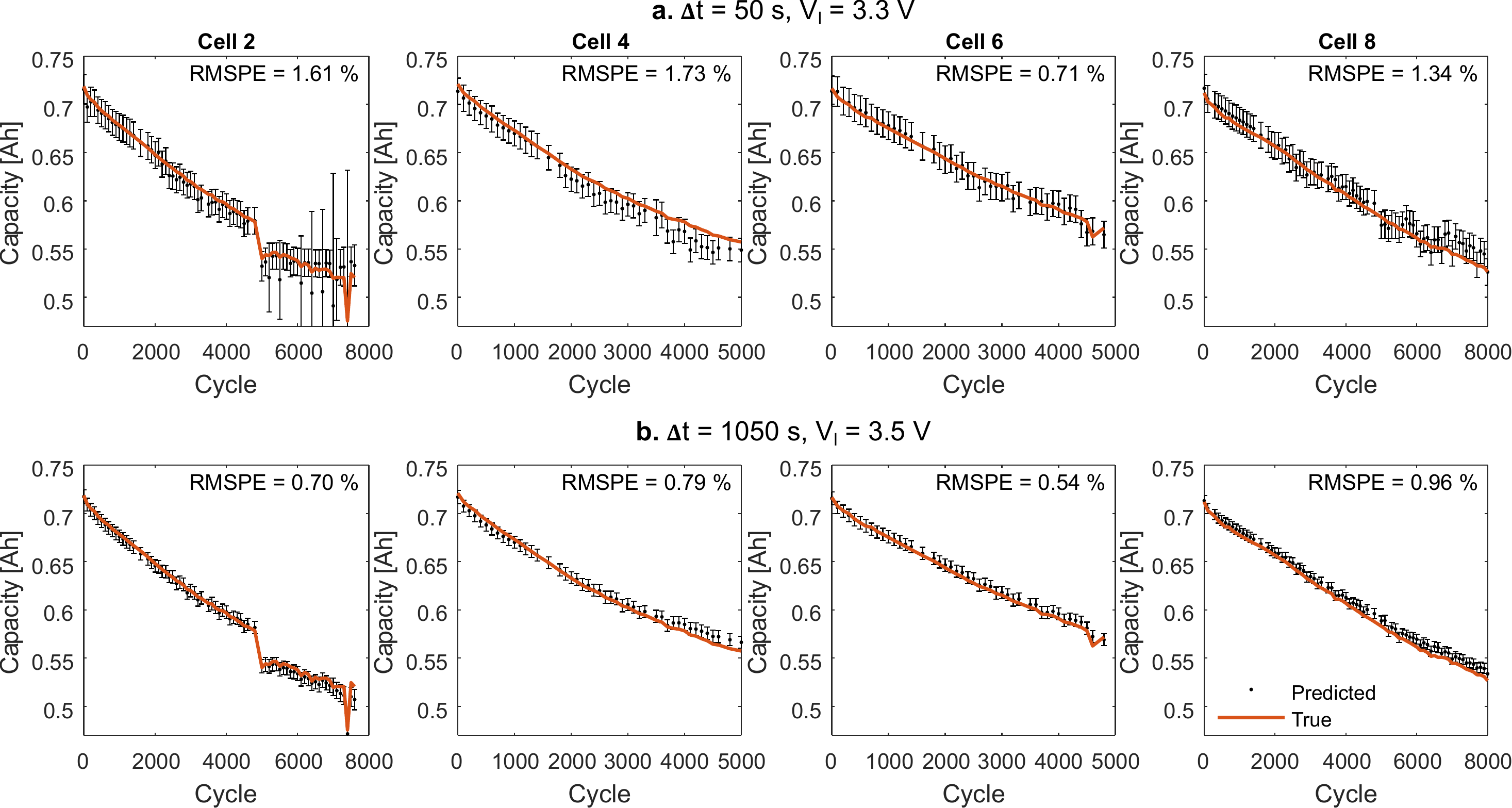}}
	\caption{\caphead{Selected results for the NASA dataset.} The red lines indicate the measured capacity and the black markers with errorbars indicate the GP-ICE estimates $\pm 2 \sigma$. \textbf{a}, Using a test duration of $\Delta t = 1050$ s and starting voltage of $V_{l} = 3.5\text{V}$, \textbf{b}, Using a test duration of $\Delta t = 10$ s and starting voltage of $V_{l} = 3.7\text{V}$.}
	\label{fig:sampleResultsNasa}
	\vspace{0.4cm}
	\vfill
	{\includegraphics[width=0.48\textwidth]{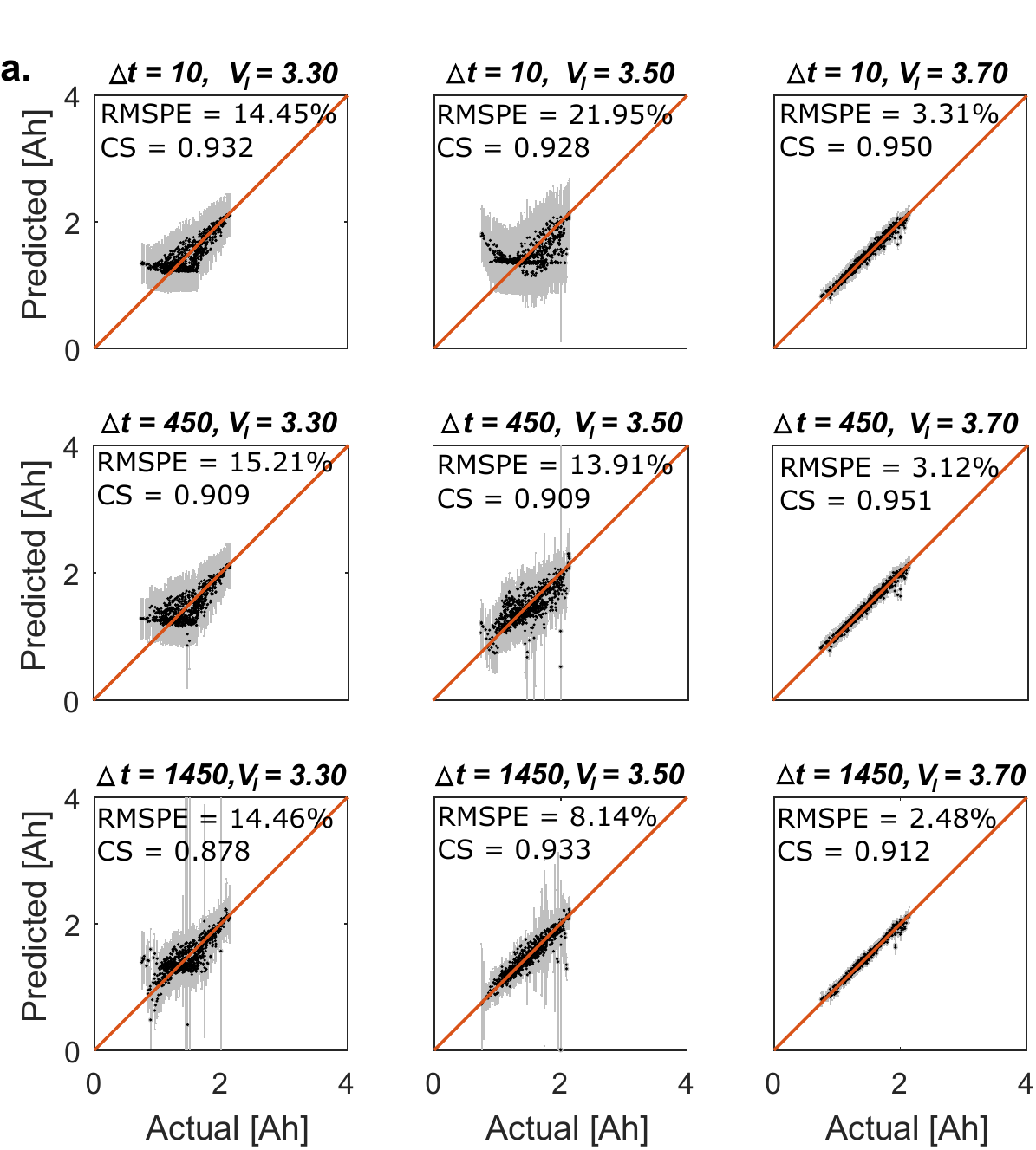}}
	\hfill
	{\includegraphics[width=0.499\textwidth]{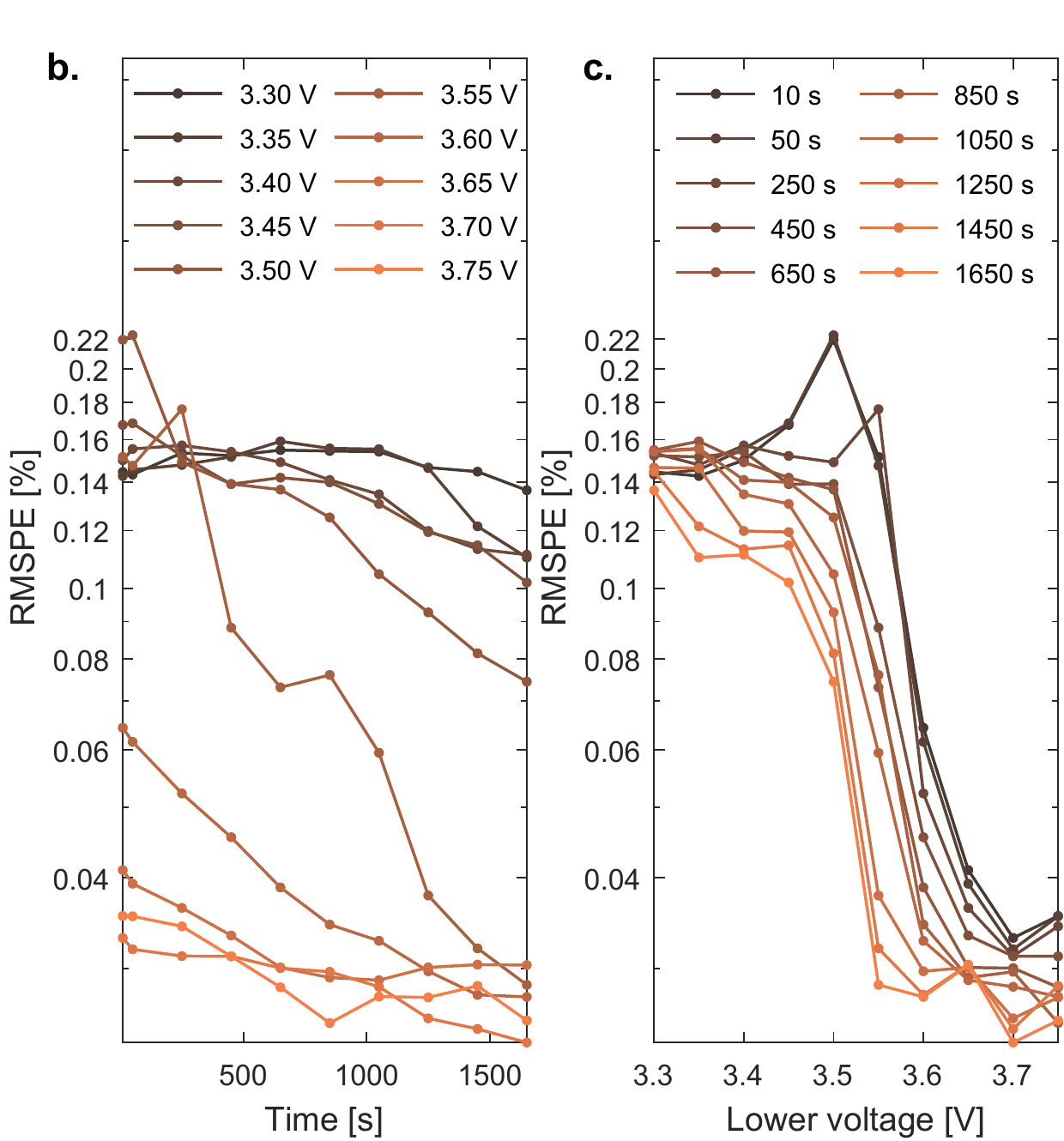}}
	\caption{\caphead{Overall results for the NASA dataset.} \rev{RMSPE} values are based on the entire dataset with each cell used once as the test set. \textbf{a}, Actual vs.\ predicted capacities for different starting voltages and measurement durations. The red line represents $y^{*} = \hat{y}^{*}$. The closer the datapoints lie to this line, the smaller the difference between the actual and predicted value. The grey lines indicate $\pm 2\sigma$ credibility intervals for each datapoint. The quoted CS values indicate the associated $\pm2\sigma$ calibration score; the closer these scores are to 0.954 the more accurate the uncertainty estimates. \textbf{b}, \rev{RMSPE} vs. measurement duration for different starting voltages. \textbf{c}, \rev{RMSPE} vs.\ starting voltage for different measurement durations. The \rev{RMSPE} generally decreases with measurement duration, but notably is also strongly affected by the starting voltage.}
	\label{fig:resultsNasa}
\end{figure*}

\subsection{NASA dataset}

Figs.\ \ref{fig:sampleResultsNasa} and \ref{fig:resultsNasa} show selected and overall results respectively for the NASA dataset, analogous to Figs.\ \ref{fig:sampleResultsOxf} and \ref{fig:resultsOxf} from the previous section.
The NASA dataset presents a greater challenge for capacity estimation since it includes cells used in 5 different cycling regimes. Moreover, even within each group the cells are not cycled with identical load profiles, but rather with statistically similar profiles generated by the same probabilistic algorithm, as discussed in Section~\ref{sec:Datasets-NASA}. Hence the GV curves used for training are more likely to differ from those used for testing than in the Oxford dataset. Nonetheless, the method performs respectably, although in general with less accuracy than for the Oxford dataset.

Fig.\ \ref{fig:sampleResultsNasa} shows results for selected cells for two combinations of $\Delta t$ and $V_{l}$.
In this case, the capacity estimates are in general less accurate than before, and the confidence intervals larger.
However, the confidence intervals do accurately reflect the model uncertainty and hence the error-bars encompass the true values in most cases.
Again, Fig.\ \ref{fig:sampleResultsNasa}b shows that surprisingly accurate estimates can be obtained with a relatively short measurement - in this case, a measurement of just 10 s duration gives accuracies of $\sim10\%$.
However, this relies on using an appropriate lower voltage -- in this case $V_{l} = 3.7$V. Indeed, the most striking aspect of these results is the strong dependence on the starting voltage, as discussed next.

Fig.\ \ref{fig:resultsNasa} shows the overall results for this dataset.
As in the previous case, increased measurement duration is shown to generally improve the capacity estimate (Fig.\ \ref{fig:resultsNasa}b).
The average calibration scores are also reasonable: CS$_{0.67\sigma} = 0.493$ and CS$_{2\sigma} = 0.920$. These are very slightly less than the true intervals, 0.5 and 0.954, indicating that the model is only slightly over-confident in its estimates.
In contrast to the previous case, the model performance is strongly dependent on the lower voltage, as shown in Fig.\ \ref{fig:resultsNasa}c (these differences in behaviour are probably attributable to the different cell chemistries of these two datasets). This figure shows that there is a cliff in the \rev{RMSPE} vs.\ $V_{l}$ curve at around 3.5 V. For starting voltages above this value, very good performance is achieved regardless of the measurement duration. This indicates that voltages in the higher range are more informative than those in the lower range for these cells.
Such insights have obvious implications for informing battery management systems on strategies for online capacity estimation.

\subsection{Comparison with IC/DV}

Lastly, GP-ICE is compared with an approach based on incremental capacity (IC) and differential voltage (DV) peak tracking.
For the latter approach, which we denote IC+DV, the location and magnitude of the largest peak in both the IC and DV curves were identified and used as inputs to the regression step. This results in 4 inputs (i.e. 2 inputs from each curve).
For the regression step the same GP model was used as for the GP-ICE method, and so any differences in performance are due to the differences in the quality of the input data (i.e. smoothed voltage data for GP-ICE vs.\ peak values of differentiated voltage data for IC+DV). Since the total number of inputs is the same as that used in the GP-ICE method, the computational requirements are identical in each case.
This IC+DV approach is similar to that used in \cite{berecibar2016online} except that in that case a neural network/SVM was used for the regression step (also, in that work, various combinations of peak features were considered, not just the most prominent peaks).
For the GP-ICE models, 6 different combinations of $\Delta t$ and $V_{l}$ were selected, and numbered as shown in Table~\ref{tab:GP-ICE-models}.
\begin{table}[!h]
	\centering
	\small
	\ra{1.1}
	\begin{tabular}{@{}lccccccc@{}}\toprule
		&  \multicolumn{7}{c}{GP-ICE}   \\
		\cmidrule{2-7}
		& 1 & 2 & 3 & 4 & 5 & 6 & \\
		\midrule
		$\Delta t$ (s) & 10 & 450 & 1,450 & 10 & 450 & 1,450 &\\
		$V_{l}$ (V) & 3.5 & 3.5 & 3.5 & 3.7 & 3.7 & 3.7 & \\
		\bottomrule
	\end{tabular}
	\caption{\caphead{GP-ICE model denotations} for 6 combinations of $\Delta t$ and $V_{l}$}
	\label{tab:GP-ICE-models}
\end{table}

\begin{figure}[hbt]
	\centering
	{\includegraphics[width=0.48\textwidth]{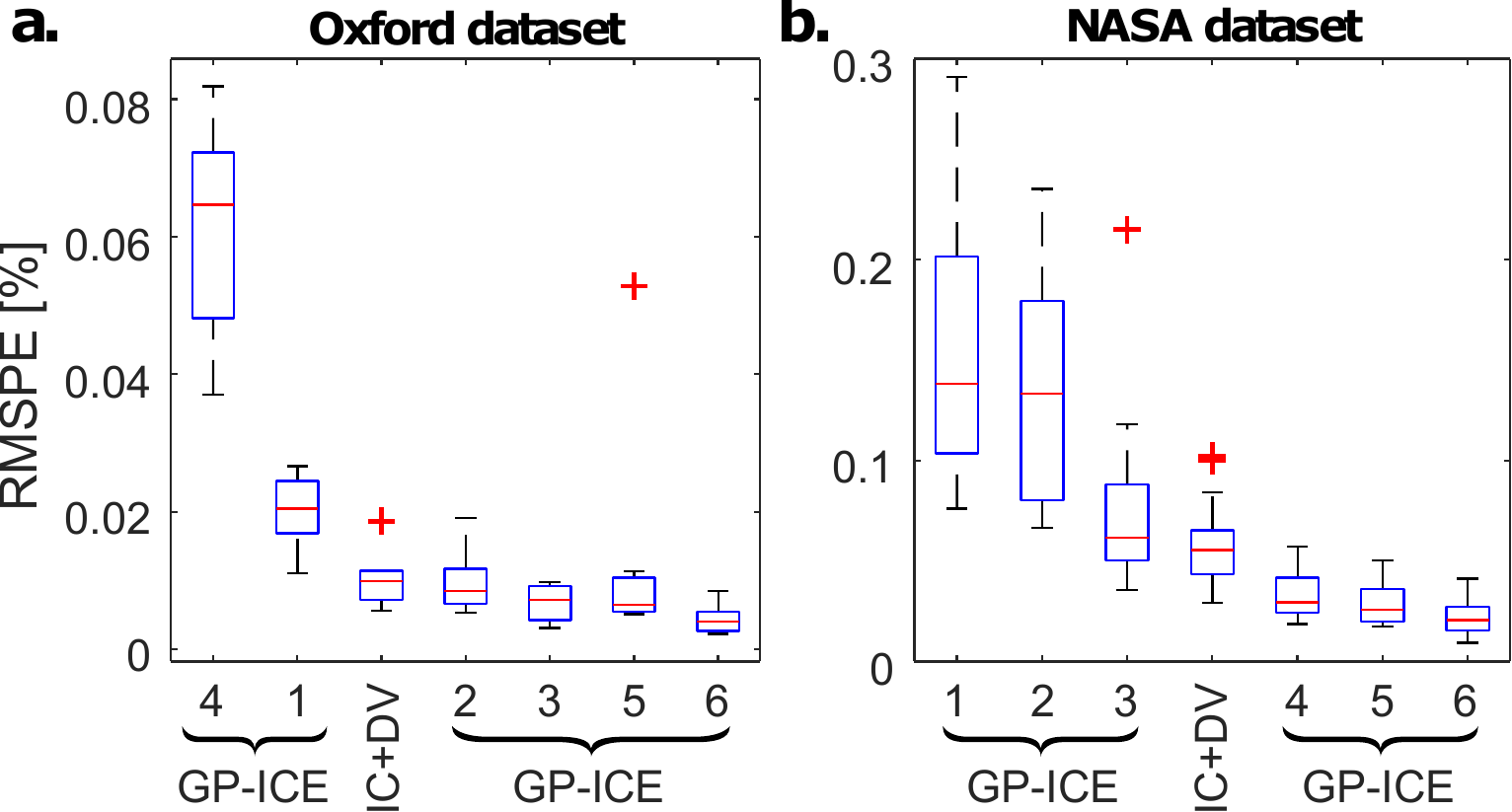}}
	\caption{\caphead{\rev{Boxplots of overall model performance}} \rev{{showing the spread in RMSPE values across all the tested cells for} \textbf{a}, Oxford dataset, \textbf{b}, NASA dataset}}
	\label{fig:rmseBoxplot}
\end{figure}

The results are shown in Fig.~\ref{fig:rmseBoxplot} and Table~\ref{tab:GP-ICE-results}.
Fig.~\ref{fig:rmseBoxplot} is a boxplot showing the spread in performance across all the tested cells, where the red lines indicate the median cell \rev{RMSPE}.
Table~\ref{tab:GP-ICE-results} shows the overall \rev{RMSPE} value when evaluated across all cells.
Bold numbers in this table indicate the best performing model for each dataset.

It is clear from these results that an appropriately selected GP-ICE test outperforms the IC+DV approach.
For the Oxford dataset, the IC+DV test is outperformed by either a 450 s GP-ICE test at $V_{l} = 3.5$V or a 1,450 s GP-ICE test at either value of $V_{l}$. In the best case ($\Delta t = 1,450$ s, $V_{l}=3.7$V), GP-ICE achieves an \rev{RMSPE} of 0.49\% compared to 1.11\% for IC+DV, a reduction by a factor of 2.26.
For the NASA dataset, IC+DV is outperformed by a test of any duration (as little as 10 s) provided the starting voltage is sufficiently high $V_{l}=3.7$V. In the best case ($\Delta t = 1,450$ s, $V_{l}=3.7$V), GP-ICE achieves an \rev{RMSPE} of 2.48\% compared to 6.55\% for IC+DV, a reduction by a factor of 2.64.

\begin{figure}[hbt]
	\centering
	\includegraphics[width=1\columnwidth]{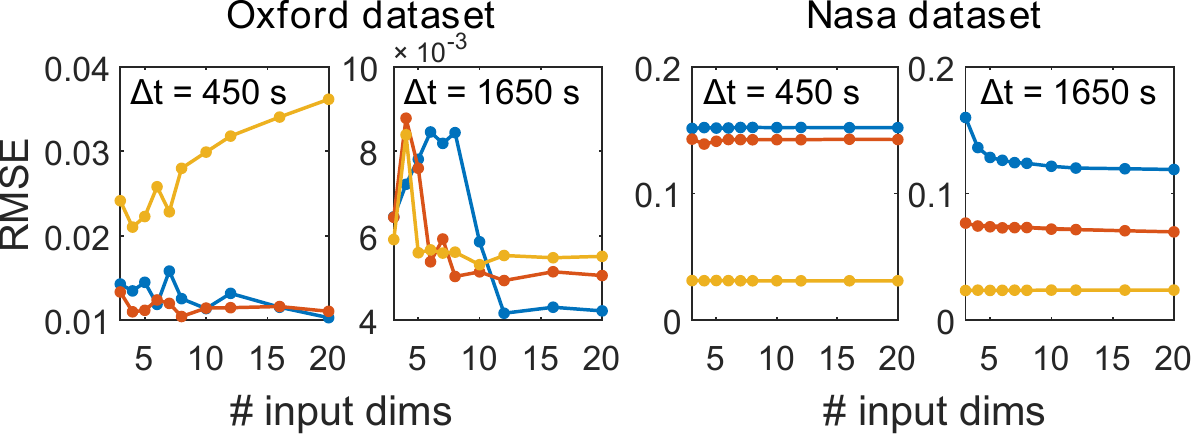}
	\caption{\textbf{Sensitivity of model accuracy to input dimensionality} for different values of $V_{l}$ and $\Delta t$. The blue, red and yellow lines indicate $V_l =$ $3.3$, $3.5$ and $3.7\,$V respectively. \textbf{a}, Oxford dataset: convergence by $n\approx4$ datapoints for $\Delta t = 450\,$s and by $n\approx10$ datapoints for $\Delta t = 1,650\,$s, \textbf{b}, NASA dataset: convergence by $n\approx4$ datapoints for both $\Delta t$ values.}
	\label{fig:sens-plot}
\end{figure}

In other cases, GP-ICE performs worse than IC+DV, most notably for lower $V_{l}$ in the NASA dataset and for shorter $\Delta t$ values in the Oxford dataset.
However, it is worth reiterating that the IC+DV approach relies on coverage of a large voltage range to capture the peaks in both the IC and DV curves, and hence these measurements could require a large and variable duration. For example, in the NASA dataset, a full GV curve takes up to 2~hrs, and so even if the peaks were separated by half this time, it would require a 1~hr test to capture both peaks.
Such a test would encompass the voltage ranges of several of the better performing \mbox{GP-ICE} tests.
Lastly, for the GP-ICE method exactly $n=4$ equispaced time samples were used as input regardless of the duration of the GV curve considered, however it is possible that the performance could be improved by increasing this number.
We tested this hypothesis with a sensitivity analysis w.r.t.\ $n$ for different values of $V_{L}$ and $\Delta t$ (Fig.~\ref{fig:sens-plot}). For $\Delta t = 450\,$s, there was negligible improvement in performance beyond $n\approx4$ inputs for either dataset.
For the Oxford dataset, minor improvements were observed up until $n \approx 10$ when $\Delta t = 1,650\,$s. Hence, some additional information could be extracted from the longer duration GV curves by increasing $n$ beyond 4. For the NASA dataset, there was little improvement beyond $n=4$ even for the longer measurement duration; this is most likely due to the lower charge rate (C/2) used for the NASA cells, meaning that even a $1,650\,$s test encompasses a relatively small voltage range.

\begin{table}[hbt]
	\centering
	\small
	\ra{1.1}
	\begin{tabular}{lc@{\hskip 0.3cm}c@{\hskip 0.15cm}c@{\hskip 0.3cm}c@{\hskip 0.3cm}c@{\hskip 0.3cm}c@{\hskip 0.3cm}c@{\hskip 0.3cm}c@{\hskip 0.3cm}}\toprule
		& \multicolumn{1}{c}{\small IC+DV} & & \multicolumn{6}{c}{GP-ICE}   \\
		\cmidrule{2-2} \cmidrule{4-9}
		& - && 1 & 2 & 3 & 4 & 5 & 6  \\
		\midrule
		\textbf{Oxford} & 1.11 && 2.10 & 1.10 & 0.74 & 6.55 & 2.10 & \textbf{0.49} \\
		\textbf{NASA} & 6.55 && 21.95 & 13.91 & 8.14 & 3.31 & 3.12 & \textbf{2.48} \\
		\bottomrule
	\end{tabular}
	\caption{\caphead{Overall model performance in \rev{RMSPE}.} The values quoted are based on the entire datasets with each cell used once as the test set. For each dataset the \rev{RMSPE} of the best performing model is shown in bold.}
	\label{tab:GP-ICE-results}
\end{table}

\section{Discussion\label{sec:discussion}}

This section briefly discusses issues related to the selection of inputs for the GP-ICE algorithm, the physical processes contributing to the observed correlations, the applicability of the approach in a practical setting. \rev{Lastly, it compares the GP-ICE approach to related work.}

\subsection{\rev{Selection of model inputs}}
Firstly, we discuss how the particular inputs to the GP-ICE algorithm -- namely time values at equispaced voltages -- were selected.
As mentioned in Section~\ref{sec:Introduction}, this choice was originally motivated by the observation that correlations existed between capacity and selected features of IC and DV curves in earlier works~\cite{weng2013board,weng2016state,berecibar2016state,berecibar2016online}. It was therefore natural to ask whether the capacity is also correlated with other portions of the curve, which do not necessarily correspond to such IC/DV peaks. 
The particular choice of inputs used in GP-ICE has a number of desirable characteristics. Firstly, by taking values spanning $V_{l}$ to $V_{u}$, the method places no restrictions on what range of voltages must be encompassed in the online test, whilst at the same time taking full advantage of whatever range it happens to include. Secondly, \textit{equispaced} measurements are expected to give the best reflection of the overall curve, for a given value of $n$.
Of course, it is possible that other design choices may improve on this performance.
In fact, the problem of estimating capacity from voltage curves could well be framed within the context of functional data analysis (FDA)~\cite{ramsay2006functional}, which is the study of information on curves or functions. 
In that case the GV curve would be treated as a functional input, and the processes of smoothing and regression would implicitly be achieved in a single principled step. An interesting area of future work would be to compare the performance of FDA against the present approach.

\subsection{\rev{Physical explanation}}
Li-ion cells undergo three primary modes of degradation: loss of lithium inventory (LLI), loss of active positive electrode material (LAM$_+$) and loss of active negative electrode material (LAM$_-$) \cite{birkl2017degradation}. These modes have observable effects on the IC/DV curves (and by extension the voltage-time data), and hence can be exploited by the GP-ICE method to infer cell capacity. Whilst elucidating the physical processes that give rise to capacity loss is an important area of study, this has been considered by several other works (e.g.\ \cite{birkl2017degradation, lewerenz2017differential}) and is therefore not the primary concern of the present paper. Rather, this work aims to highlight that raw voltage measurements can be used to infer the capacity without necessarily knowing the exact mechanisms through which this occurs. This is in fact core to the advantage of GP-ICE: since it does not rely on cell specific knowledge such as the expected locations and numbers of peaks in IC/DV curves, it could be directly applied to other cell chemistries without modification.
Of course, there is no guarantee of equivalent results to those obtained here -- the performance is dependent on how strongly the galvanostatic voltage-time data are correlated with the cell capacity, something which may vary from cell to cell and across voltage ranges, as the earlier results show.
However, the important point is that there is no need to encode any cell-specific information in our model -- the capacity estimation is achieved automatically in any case.
This generality also opens up the possibility of applying the method to portions of constant-current data within otherwise dynamic drive cycles.
This is likely to be non-trivial due to dynamics in the cell; however, if long enough portions of constant current are available, then it may give satisfactory results.

\subsection{\rev{Practical application}}
There are many practical scenarios in which GP-ICE could be applicable. For instance, in EV applications, the vast majority of charging stations output a power of less than $22$ kW~\cite{sbordone2015ev}, which would equate to $<0.5$~C for a typical EV battery pack.
\rev{Nonetheless, the effect of C-rate on performance could be considered in future work to establish whether the method would be feasible using higher power charging/discharging. It is probable that higher pre-specified C-rates may result in lower performance -- since higher C-rates result in some of the subtler features of the OCV curve being “smoothed out” by the cell impedance -- but it is not clear to what extent this would be the case.}
\rev{Another important consideration} is the application of the technique under variable ambient temperature conditions.
\rev{The present results apply to a single temperature for each dataset; however, variations in temperature can result in significant changes to the measured impedance and OCV~\cite{richardson2015sensorless} and so accounting for this variation will be essential for the method to be applied in different ambient temperature conditions.
This could be achieved provided appropriate training data are available encompassing the relevant range of temperatures.
We emphasise that this would not necessarily require a large increase in experimental effort: for instance, to include additional temperatures it is merely necessary to repeat the reference charge/discharge measurement step under each of the required temperatures for each cell. The most time intensive portion of the test -- namely ageing the cells by repeated operation under various drive cycles -- could remain unchanged. Also, it should be noted that these limitations apply equally to a number of other approaches to capacity estimation, including Incremental Capacity and Differential Voltage analysis.}

\subsection{\rev{Related work}}

\rev{Lastly, we briefly compare the present approach with other recent studies related to feature extraction of online measurements for battery SOH estimation. We consider here only the most relevant studies; the reader is referred to the review studies \cite{farmann2015critical,zhang2011review,ungurean2017battery} for details of other approaches.}

\rev{You et al.\ \cite{you2017diagnosis} presented an approach which uses a Recurrent Neural Network trained on partial charge curves for estimating cell capacity. This is similar to our approach but with some key differences. Specifically, our GP-ICE approach: (i) employs a Gaussian process method for the regression step, which provides confidence in the capacity estimates, (ii) uses Savitzky-Golay filtering as a preprocessing step to improve signal to noise ratio, (iii) selects a subset of the smoothed data in order to minimise computational overhead -- this is a necessary requirement given the higher computational overhead of GPs compared to neural networks. Moreover, our method shows how the performance of the capacity estimates varies as a function of the starting voltage and measurement duration, something which has not been demonstrated in previous work.
On the other hand, the method of \cite{you2017diagnosis} exploits the sequential nature of the charge curves, unlike our approach, which ignores any correlation between the inputs. An interesting area of future work could involve accounting for correlations between the inputs by encoding recurrent behaviour into the kernel of the GP function (such as in the method presented in~\cite{al2016learning}) in order to achieve the benefits of both of these approaches.
}

\rev{
Differential Thermal Voltammetry (DTV) is another approach to capacity estimation that has been introduced very recently~\cite{wu2015differential, merla2016novel, shibagaki2018tracking}.
DTV tracks battery degradation through phase transitions, and the resulting entropic heat, occurring in the electrodes, by means of temperature vs.\ time measurements under relatively high current loads. In some respects, DTV is similar to Differential Voltage Analysis but using temperature, rather than voltage, measurements. The key advantage over Differential Voltage Analysis is that DTV is applicable using higher currents and hence enables shorter diagnostic tests.
DTV could in fact be \textit{complementary} to the GP-ICE approach presented here: e.g.\ GP-ICE could be applied using measurements of temperature rather than voltage, combining the advantages of both approaches.
}


\FloatBarrier
\section{Conclusions\label{sec-conclusions}}

This paper has introduced GP-ICE, a technique for estimating battery capacity using small portions of voltage-time data under constant current (galvanostatic) operation.
The primary novel aspects of our approach are as follows:
\begin{enumerate}
	\item \textbf{Operates on raw voltage data:}
	GP-ICE dispenses with the interpretation of galvanostatic voltage (GV) data as incremental capacity or differential voltage curves, and instead involves directly performing regression using the voltage/time data as inputs. 
	\item \textbf{Automatic input extraction:}
	To enable automatic identification of inputs for a new cell, GP-ICE uses a two-step process of (i) smoothing the voltage data and (ii) sampling voltages from the smoothed data to obtain the inputs to the regression model.
	\item \textbf{Bayesian non-parametric regression}: GP-ICE utilises a probabilistic paradigm, unlike previous works. It therefore adapts to the complexity of the data and avoids over-fitting, whilst also providing accurate estimates of uncertainty in its predictions
\end{enumerate}
Features (1) and (2) above have a number of benefits, including mitigating the inaccuracy introduced by differentiating the voltage-time data, enabling capacity estimates using arbitrary portions of the voltage curve, and overcoming the need for cumbersome analysis of the voltage-time data for a new cell to identify the features of interest.
Feature (3) is also important: through the use of a Bayesian non-parametric regression technique, Gaussian processes regression, the model adapts to the complexity of the data and avoids over-fitting.

Concretely, GP-ICE was shown to outperform IC/DV peak tracking by a factor $\sim$2.5 in terms of \rev{RMSPE}, whilst also providing the various aforementioned advantages such as greater flexibility, shorter diagnostic test requirements, and the provision of accurate estimates of uncertainty in its predictions.
It also provides insight into which voltage ranges are most informative, and hence may inform a BMS as to when best to perform a diagnostic test.

Future work should consider accounting for \rev{variable ambient temperatures and/or higher pre-specified C-rates -- this should be feasible provided training data under the relevant temperatures/C-rates are acquired during each reference charge/discharge step during the ageing experiments.}


%
%
%

\appendix[Gaussian process regression]

A Gaussian process (GP)~\cite{rasmussen2006gaussian} defines a probability distribution over functions, and is denoted as:
\begin{equation}
f(\mathbf{x}) \sim \mathcal{GP}\bigl(m(\mathbf{x}), \kappa(\mathbf{x},\mathbf{x}')\bigr),
\end{equation}
where $m(\mathbf{x})$ and $\kappa(\mathbf{x},\mathbf{x}')$ are the mean and covariance functions respectively, denoted by
\begin{align}
m(\mathbf{x}) & =
\mathbb{E} [f(\mathbf{x})], \\
\kappa(\mathbf{x}, \mathbf{x}') & =
\mathbb{E} [\left(f(\mathbf{x}) - m(\mathbf{x})\right) \left(f(\mathbf{x}') - m(\mathbf{x}')\right)^T].
\end{align}
For any finite collection of input points, say ${X} = \mathbf{x}_1,...,\mathbf{x}_{N_D}$, this process defines a probability distribution $p\left( f(\mathbf{x}_1),...,f(\mathbf{x}_{N_D}) \right)$ that is jointly Gaussian, with some mean $\mathbf{m} (\mathbf{x})$ and covariance $\mathbf{K} (\mathbf{x})$ given by $K_{ij} = \kappa(\mathbf{x}_i,\mathbf{x}_j)$.

Gaussian process regression is a way to achieve non-parametric regression with Gaussian processes.
The key idea is that, rather than postulating a parametric form for the function $f({\mathbf{x}, \phi})$ and estimating the parameters $\phi$ (as in parametric regression), we instead assume that the function $f(\mathbf{x})$ is a sample from a Gaussian process as defined above.

In this work, we use the Mat\'ern covariance function:
\begin{equation}
\kappa_{\text{Ma}}({x}-{x}') = 
\sigma_f^2 \frac{2^{1-\nu}}{\Gamma(\nu)}
\left(\sqrt{2\nu}\frac{({x}-{x}')}{\rho}\right)^{\nu}
\mathcal{R}_{\nu}\left(\sqrt{2\nu}\frac{({x}-{x}')}{\rho}\right)
,
\end{equation}
with smoothness hyperparameter, $\nu=5/2$ (larger $\nu$ implies smoother functions)
and $\mathcal{R}_{\nu}$ is the modified Bessel function.
This kernel was chosen as it suitable for functions with varying degrees of smoothness, although similar performance was observed using other common kernels, including the Squared Exponential~\cite{rasmussen2006gaussian}.
The mean function is commonly defined as $m(\mathbf{x})=0$, and for convenience this convention is followed here.

Now, if we observe a labelled training set of input-output pairs
$\mathcal{D} = \{(\mathbf{x}_i, {y}_i)\}_{i=1}^{N_D}$, predictions can be made at test indices ${X}^*$ by computing the conditional distribution $p(\mathbf{y}^* \vert {X}^*, {X}, \mathbf{y})$.
This can be obtained analytically by the standard rules for conditioning  Gaussians~\cite{murphy2012machine}, and (assuming a zero mean for notational simplicity) results in a Gaussian distribution given by:
\begin{equation}
p(\mathbf{y}^* \vert {X}^*, {X}, \mathbf{y}) = \mathcal{N}(\mathbf{y}^* \vert \mathbf{m}^*, \mathbf{\sigma}^*)
\end{equation}
where
\begin{align}
\mathbf{m}^* & = \mathbf{K}({X}, {X}^*)^T
\mathbf{K}({X}, {X})^{-1}
\mathbf{y}\\
\mathbf{\sigma}^* & = \mathbf{K}({X}^*,{X}^*)
- \mathbf{K}({X}, {X}^*)^T
\mathbf{K}({X}, {X})^{-1}
\mathbf{K}({X}, {X}^*).
\end{align}

The values of the covariance hyperparameters $\theta= \{\sigma_f, \rho\}$ may be optimised by minimising the negative log marginal likelihood defined as $\text{NLML} = -\log p(\mathbf{y} \vert {X}, \theta)$.
Minimising the NLML automatically performs a trade-off between bias and variance, and hence ameliorates over-fitting to the data.		
Given an expression for the NLML and its derivative w.r.t $\theta$ (both of which can be obtained in closed form), the value of $\theta$ can be estimated using any standard gradient-based optimizer. In the present case, the GPML toolbox~\cite{rasmussen2010gpml} implementation of conjugate gradients was used.

\section{Supplementary material}

\begin{figure*}[hbt]
	\centering
	\includegraphics[width=0.75\textwidth]{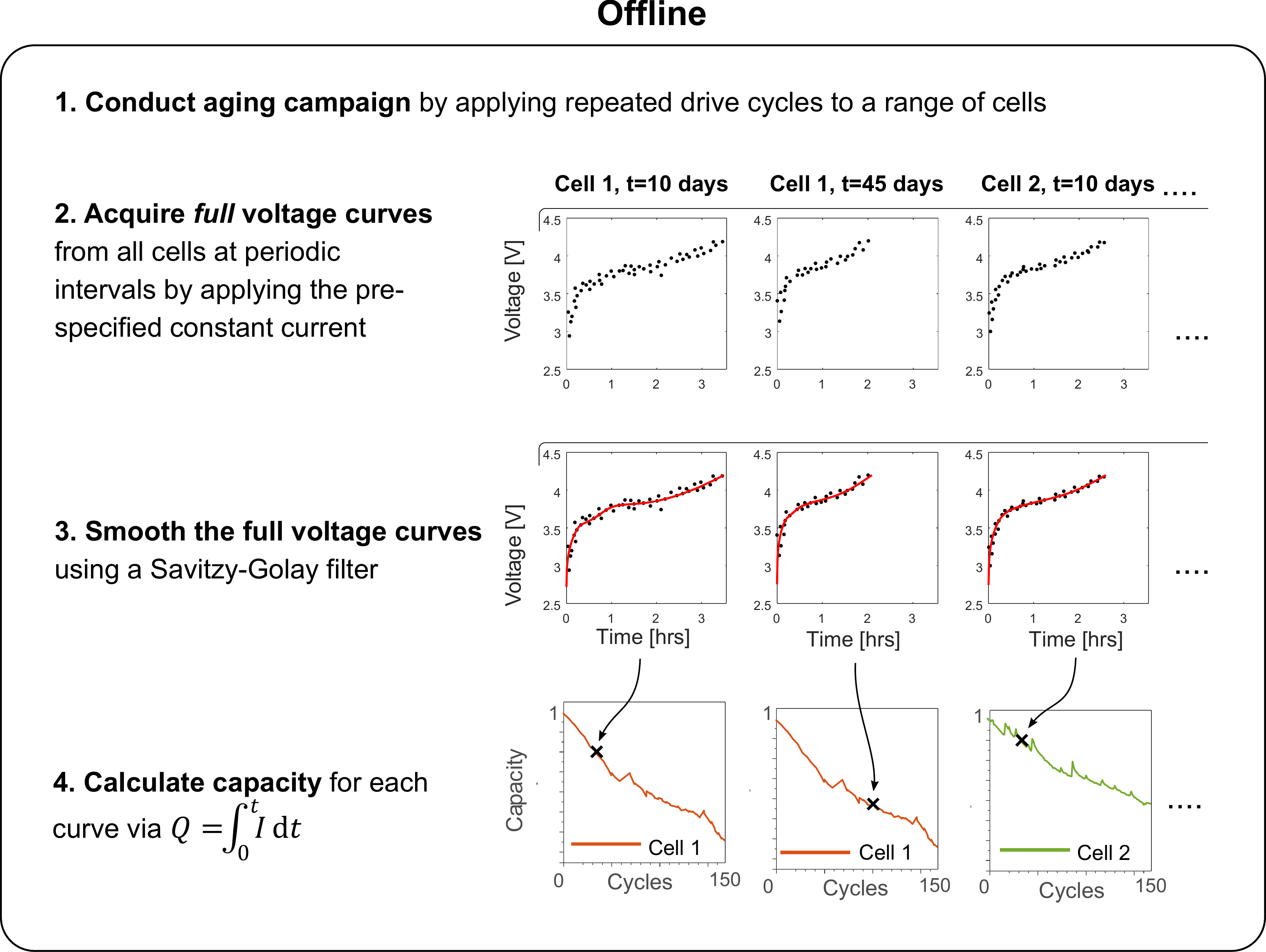}
	
	\vspace{0.5cm}
	
	\includegraphics[width=0.9\textwidth]{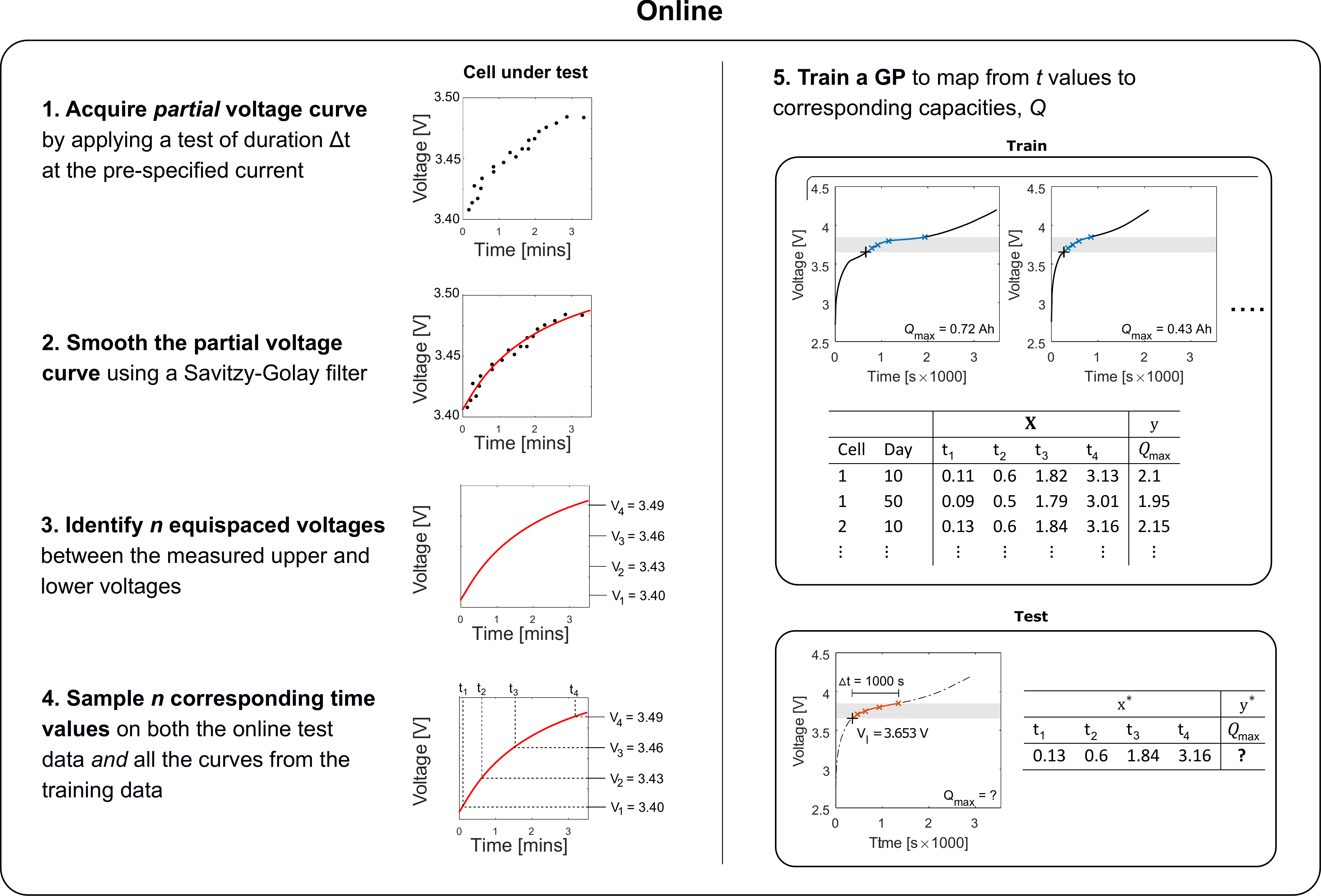}
	\caption{\textbf{GP-ICE flow diagram}. Note that the data used in these plots was generated for illustration purposes. See Section II.A for further details.}
	
	\label{fig:schematic}
\end{figure*}

\section*{Acknowledgement}
This work was funded by an RCUK Engineering and Physical Sciences Research Council grant, ref. EP/K002252/1.

\ifCLASSOPTIONcaptionsoff
  \newpage
\fi



\bibliographystyle{IEEEtran}
\bibliography{GPICE_IEEE_references}

\end{document}